\documentclass[dvips]{acta}
\usepackage{supertabular,lscape,epsfig}
\usepackage{amssymb}
\usepackage{amsmath}
\DeclareSymbolFont{ppa}{OT1}{ppl}{m}{it}
\DeclareMathSymbol{\vv}{\mathalpha}{ppa}{'166}

\newfont{\hb}{rphvb at 10pt}
\newfont{\hbo}{rphvbo at 10pt}
\newfont{\bitt}{rptmbi at 12pt}
\newfont{\bits}{rptmbi at 11pt}

\SetPages{1}{1}

\SetVol{64}{2014}

\begin{document}

\newcommand{\TabCapp}[2]{\begin{center}\parbox[t]{#1}{\centerline{
  \small {\spaceskip 2pt plus 1pt minus 1pt T a b l e}
  \refstepcounter{table}\thetable}
  \vskip2mm
  \centerline{\footnotesize #2}}
  \vskip3mm
\end{center}}

\newcommand{\TTabCap}[3]{\begin{center}\parbox[t]{#1}{\centerline{
  \small {\spaceskip 2pt plus 1pt minus 1pt T a b l e}
  \refstepcounter{table}\thetable}
  \vskip2mm
  \centerline{\footnotesize #2}
  \centerline{\footnotesize #3}}
  \vskip1mm
\end{center}}

\newcommand{\MakeTableSepp}[4]{\begin{table}[p]\TabCapp{#2}{#3}
  \begin{center} \TableFont \begin{tabular}{#1} #4
  \end{tabular}\end{center}\end{table}}

\newcommand{\MakeTableee}[4]{\begin{table}[htb]\TabCapp{#2}{#3}
  \begin{center} \TableFont \begin{tabular}{#1} #4
  \end{tabular}\end{center}\end{table}}

\newcommand{\MakeTablee}[5]{\begin{table}[htb]\TTabCap{#2}{#3}{#4}
  \begin{center} \TableFont \begin{tabular}{#1} #5
  \end{tabular}\end{center}\end{table}}

\newfont{\bb}{ptmbi8t at 12pt}
\newfont{\bbb}{cmbxti10}
\newfont{\bbbb}{cmbxti10 at 9pt}
\newcommand{\uprule}{\rule{0pt}{2.5ex}}
\newcommand{\douprule}{\rule[-2ex]{0pt}{4.5ex}}
\newcommand{\dorule}{\rule[-2ex]{0pt}{2ex}}
\def\thefootnote{\fnsymbol{footnote}}
\begin{Titlepage}
\Title{OGLE-BLG-RRLYR-12245: An RR Lyrae Star that Switched\\
from a Double- to Single-mode Pulsation\footnote{Based on observations
obtained with the 1.3-m Warsaw telescope at the Las Campanas Observatory
of the Carnegie Institution for Science.}}
\Author{I.~~S~o~s~z~y~ñ~s~k~i$^1$,~~
W.\,A.~~D~z~i~e~m~b~o~w~s~k~i$^1$,~~
A.~~U~d~a~l~s~k~i$^1$,\\
M.\,K.~~S~z~y~m~a~ñ~s~k~i$^1$,~~
M.~~K~u~b~i~a~k$^1$,~~
G.~~P~i~e~t~r~z~y~ñ~s~k~i$^{1,2}$,\\
\L.~~W~y~r~z~y~k~o~w~s~k~i$^{1,3}$,~~
K.~~U~l~a~c~z~y~k$^1$,~~
R.~~P~o~l~e~s~k~i$^{1,4}$,~~
S.~~K~o~z~³~o~w~s~k~i$^1$,\\
P.~~P~i~e~t~r~u~k~o~w~i~c~z$^1$,~~
J.~~S~k~o~w~r~o~n$^1$,~~
and~~ P.~~M~r~ó~z$^1$}
{$^1$Warsaw University Observatory, Al.~Ujazdowskie~4, 00-478~Warszawa, Poland\\
e-mail: (soszynsk,wd,udalski)@astrouw.edu.pl\\
$^2$ Universidad de Concepción, Departamento de Astronomia, Casilla 160--C, Concepción, Chile\\
$^3$ Institute of Astronomy, University of Cambridge, Madingley Road, Cambridge CB3 0HA, UK\\
$^4$ Department of Astronomy, Ohio State University, 140 W. 18th Ave., Columbus, OH 43210, USA}
\Received{~}
\end{Titlepage}
\Abstract{We report the discovery of an RR~Lyrae star that experienced a
switching of its pulsation mode. OGLE-BLG-RRLYR-12245 was discovered as a
double-mode RRd star from the observations conducted in years 2001-2006 during the
third phase of the Optical Gravitational Lensing Experiment (OGLE-III). The
OGLE-IV observations carried out since 2010 reveal that this object is now
a fundamental-mode RRab star, with no sign of the first-overtone
pulsation. The analysis of the OGLE photometry shows that the final stage
of the mode switching occurred on a relatively short timescale of a few
months in 2005. We study the behavior of the star during this process,
showing changes of the pulsational amplitudes and periods. We also discuss
possible causes for the mode switching in RR~Lyr stars.}
{Stars: variables: RR~Lyrae -- Stars: oscillations -- Stars: Population II
-- Stars: horizontal-branch}

\Section{Introduction}
Mode switching in RR~Lyr stars is a very rarely observed phenomenon. So
far, only one RR~Lyr variable that changed its pulsation mode has been
detected. Variable star V79 in the globular cluster M3 was known for
decades as a fundamental-mode RR~Lyr star (RRab), however in 1992 V79
became a double-mode pulsator (RRd) with the first-overtone mode dominating
(Clement \etal 1997, Kaluzny \etal 1998, Clement and Goranskij 1999). Then,
in 2007 this star again changed its pulsation mode and returned to the sole
fundamental mode (Goranskij 2010). The mode switches in V79 were
accompanied with significant changes of the pulsational period.

In this paper, we report the discovery of another RR~Lyr star that recently
switched its pulsation modes. OGLE-BLG-RRLYR-12245 was identified in the
Galactic bulge by Soszyñski \etal (2011) as an RRd star on the basis of the
observations collected in years 2001-2006 during the third phase of the Optical
Gravitational Lensing Experiment (OGLE-III). Double-mode RR~Lyr stars are
very rare in the Galactic bulge. Among $16\;836$ RR~Lyr stars detected in
the OGLE-III fields toward the Galactic center only
91 objects (0.5\% of the total sample) were classified as RRd stars
(Soszyñski \etal 2011).

\begin{figure}[t]
\includegraphics[width=12.7cm]{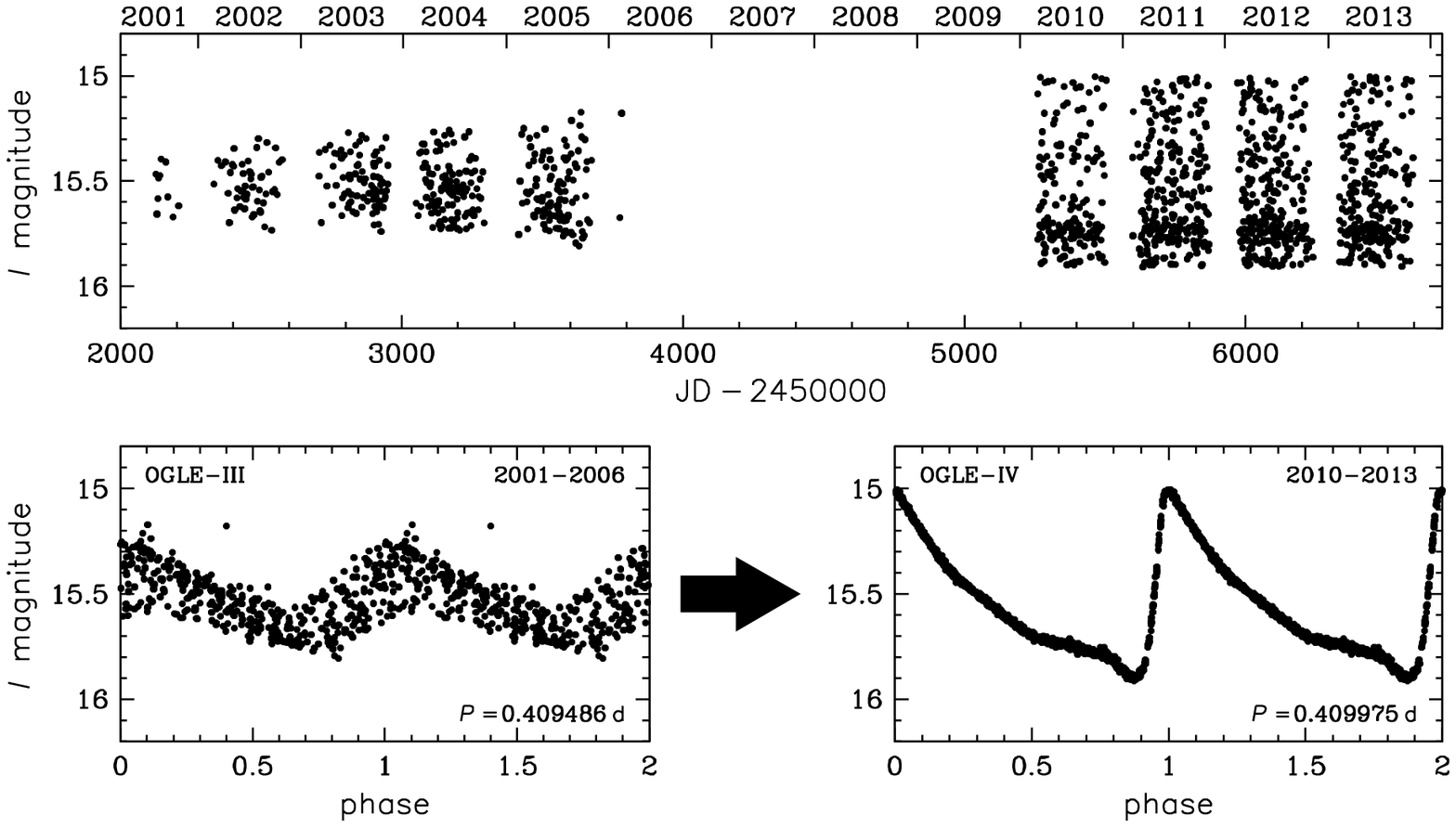}
\FigCap{{\it Upper panel}: unfolded {\it I}-band light curve of
OGLE-BLG-RRLYR-12245 obtained by the OGLE survey in the years
2001-2013. {\it Lower panels}: OGLE-III ({\it left panel}) and OGLE-IV
({\it right panel}) light curves of OGLE-BLG-RRLYR-12245 folded with the
fundamental-mode periods. Note the difference in the periods before and
after the mode switching.}
\end{figure}

Recently, we searched for RR~Lyr variables in the OGLE-IV bulge fields,
photometrically monitored since 2010, and we significantly increased the
number of known stars of this type in the central regions of the Milky
Way. This new collection of RR~Lyr variables and other pulsating stars will
be presented in forthcoming papers. We also inspected the OGLE-IV light
curves of the already known RR~Lyr stars and noticed that
OGLE-BLG-RRLYR-12245 drastically changed its pulsational properties. This
metamorphosis is clearly visible in Fig.~1 presenting the OGLE-III and
OGLE-IV light curves of our target folded with the fundamental-mode
periods. Observations collected in the years 2010-2013 show that
OGLE-BLG-RRLYR-12245 is currently a typical RRab star, with no sign of the
first-overtone period or any other secondary variability, including the
Blazhko modulation. The period of the fundamental-mode oscillations
increased by 0.000489~d (0.12\%) between 2005 and 2010. Table~1 provides
information about the identification, position and brightness of
OGLE-BLG-RRLYR-12245.

\MakeTable{l@{\hspace{5mm}}l}{12.5cm}{Properties of OGLE-BLG-RRLYR-12245}
{\hline
\noalign{\vskip3pt}
OGLE-III field and star's number & BLG175.1~~144525 \\
OGLE-IV field and star's number  & BLG514.03~~97374 \\
Right ascension (J2000)          & 18\uph03\upm54\zdot\ups05 \\
Declination (J2000)              & $-31\arcd25\arcm10\zdot\arcs8$ \\
Galactic longitude               & $359\zdot\arcd846102$ \\
Galactic latitude                & $-4\zdot\arcd663256$ \\
Mean apperent luminosity $\langle{I}\rangle$ & 15.525 mag \\
Mean apperent luminosity $\langle{V}\rangle$ & 17.277 mag \\
\noalign{\vskip3pt}
\hline}

\Section{Observations and Data Analysis}

All the photometric data discussed in this paper were obtained with the
1.3-m Warsaw Telescope located at Las Campanas Observatory, Chile. The
observatory is operated by the Carnegie Institution for Science. The OGLE
survey began regular photometric monitoring of the stellar field around
OGLE-BLG-RRLYR-12245 at the beginning of the OGLE-III phase, in August
2001. The field was continuously observed (with the seasonal breaks) till
February 2006. In total, 380 observing points in the Cousins {\it I}-band
filter and a few measurements in the Johnson {\it V}-band were
collected. Observations of these regions were resumed at the beginning of
the OGLE-IV project, in March 2010, and are continued to this day with 959
{\it I}-band measurements already collected. Detailed descriptions of the
instrumentation, photometric reductions and astrometric calibrations of the
OGLE data are available in Udalski (2003) and Udalski \etal (2008).

\begin{figure}[t]
\includegraphics[width=12.7cm]{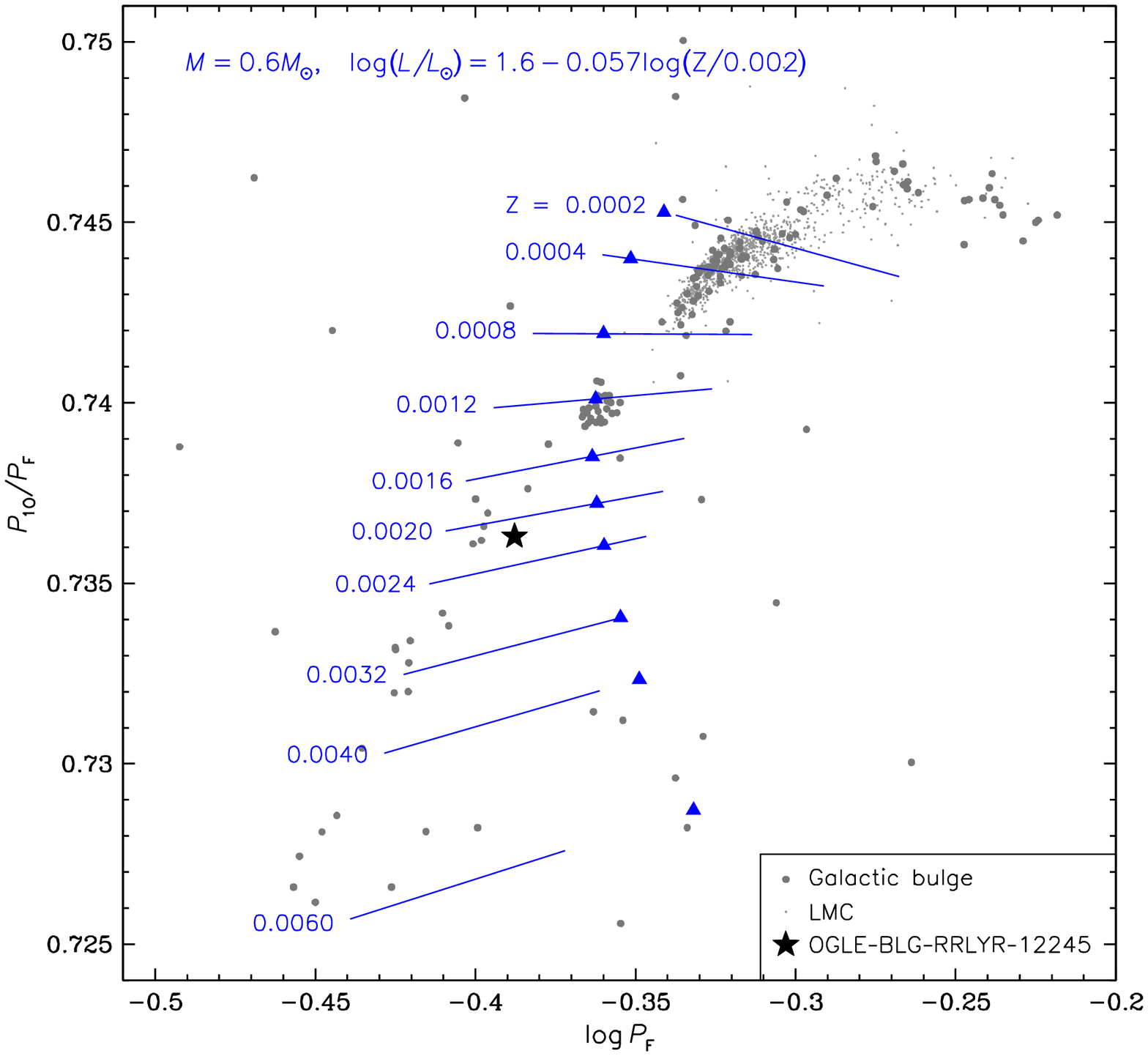}
\FigCap{Petersen diagram for RRd stars in the Galactic bulge (large grey
circles) and LMC (small grey dots). Large black star shows the position of
OGLE-BLG-RRLYR-12245. The blue lines represent calculated period ratios for
selected models covering the instability strip. The metal abundance
parameters, $Z$, are given next to each model. The blue triangles show the
loci of the $2/P_{\rm 1O}=1/P_{\rm F}+1/P_{\rm 2O}$ resonance. Theoretical
periods were calculated for the envelope models with $M=0.6M_\odot$, and
luminosity in solar units satisfying the $\log L=1.6-0.057\log(Z/0.002)$
relation. The effective temperature at the blue edge is $\log T_{\rm
eff}=3.853-0.19(\log L-1.6)$ and at the red edge is by 0.02~dex lower.}
\end{figure}

In the OGLE-III catalog of RR~Lyr stars in the Galactic bulge (Soszyñski
\etal 2011), OGLE-BLG-RRLYR-12245 was classified as an RRd star. Its
position in the Petersen diagram (period ratios versus logarithm of the
longer periods) is shown in Fig.~2, in which we included also other RRd
stars detected by OGLE-III and OGLE-IV in the Galactic bulge and Large
Magellanic Cloud (LMC, Soszyñski \etal 2009). As can be seen in Fig.~2, the
first-overtone/fundamental-mode period ratio in OGLE-BLG-RRLYR-12245 is
quite typical for RRd stars in the Galactic bulge, although it is smaller than
period ratios observed in RRd stars from the LMC and other studied galaxies
and clusters. Such low period ratios in the bulge double-mode RR~Lyr stars
are likely to be due to their exceptionally high metallicities.
According to the linear pulsation models similar to those used in
Soszyñski \etal (2011), OGLE-BLG-RRLYR-12245 has metal abundance
parameter $Z$ between 0.0020 and 0.0024. The exact value depends
weakly on the adopted mass.

\begin{figure}[t]
\includegraphics[width=12.7cm]{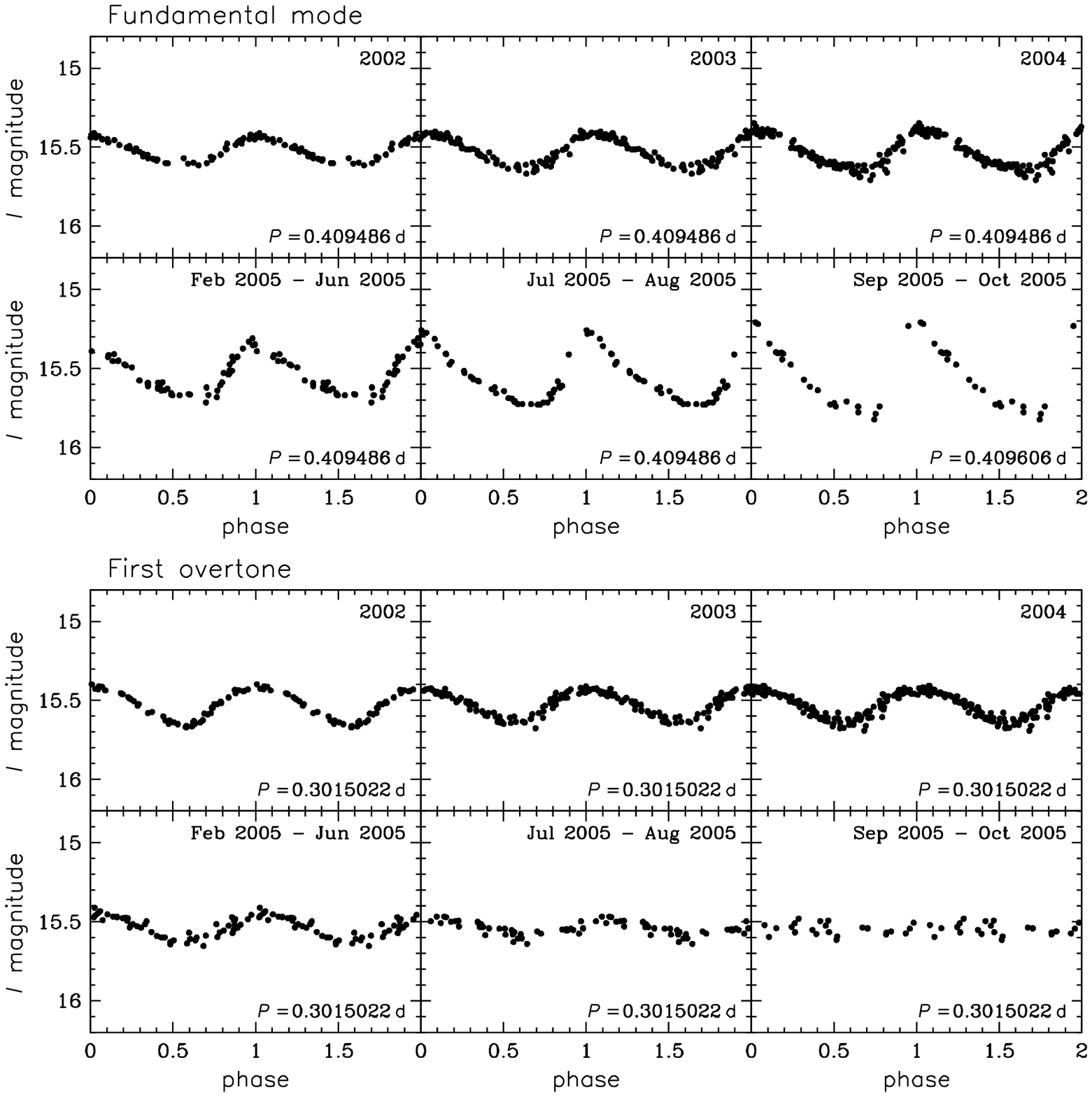}
\FigCap{Evolution of the fundamental-mode ({\it upper panels}) and
first-overtone ({\it lower panels}) {\it I}-band light curves of
OGLE-BLG-RRLYR-12245. Both pulsation modes have been separated with the
Fourier techniques. First three panels in each mode present light curves
obtained within the years 2002, 2003, and 2004, respectively. The remaining
panels show observations collected in 2005 -- from left to right: between
February and June, from July to August, and from September to October 2005.}
\end{figure}

The morphology of the studied light curve substantially changed during
4.5~years of the OGLE-III monitoring. To investigate this behavior, we
divided the light curve into parts covering shorter time spans. In each
part, the fundamental- and first-overtone modes were separated by fitting
of the double-frequency Fourier sum. The periods of the modes were also
optimized during the fitting. The resulting light curves obtained in the
years 2002-2005 are presented in Fig.~3. Light curves obtained in 2001 and
2006 were omitted due to a small number of measurements collected in these
years.

\vspace*{9pt}
\Section{Period and Amplitude Changes of OGLE-BLG-RRLYR-12245}

Fig.~3 illustrates the process of the mode switching in
OGLE-BLG-RRLYR-12245. In 2002, our target was a double-mode pulsator with
the first overtone as the dominant mode. In the subsequent two years, the
fundamental-mode oscillations increased in strength, while the amplitude of
the first-overtone mode decreased in 2003 and slightly increased in
2004. In the years 2002-2004 both pulsation periods remained constant
within the uncertainties.

The phenomenon of mode switching accelerated in 2005, thus the light curve
obtained in that year was divided into three parts. The time spans of
each part are provided in different panels of Fig.~3. One can see a rapid
increase of the fundamental-mode amplitude and decrease of the
first-overtone amplitude. By the end of 2005, the first-overtone mode
disappeared completely. At the same time, the fundamental-mode period
noticeably changed, namely increased from 0.409486~d to 0.409606~d.

\begin{figure}[t]
\includegraphics[width=12.7cm]{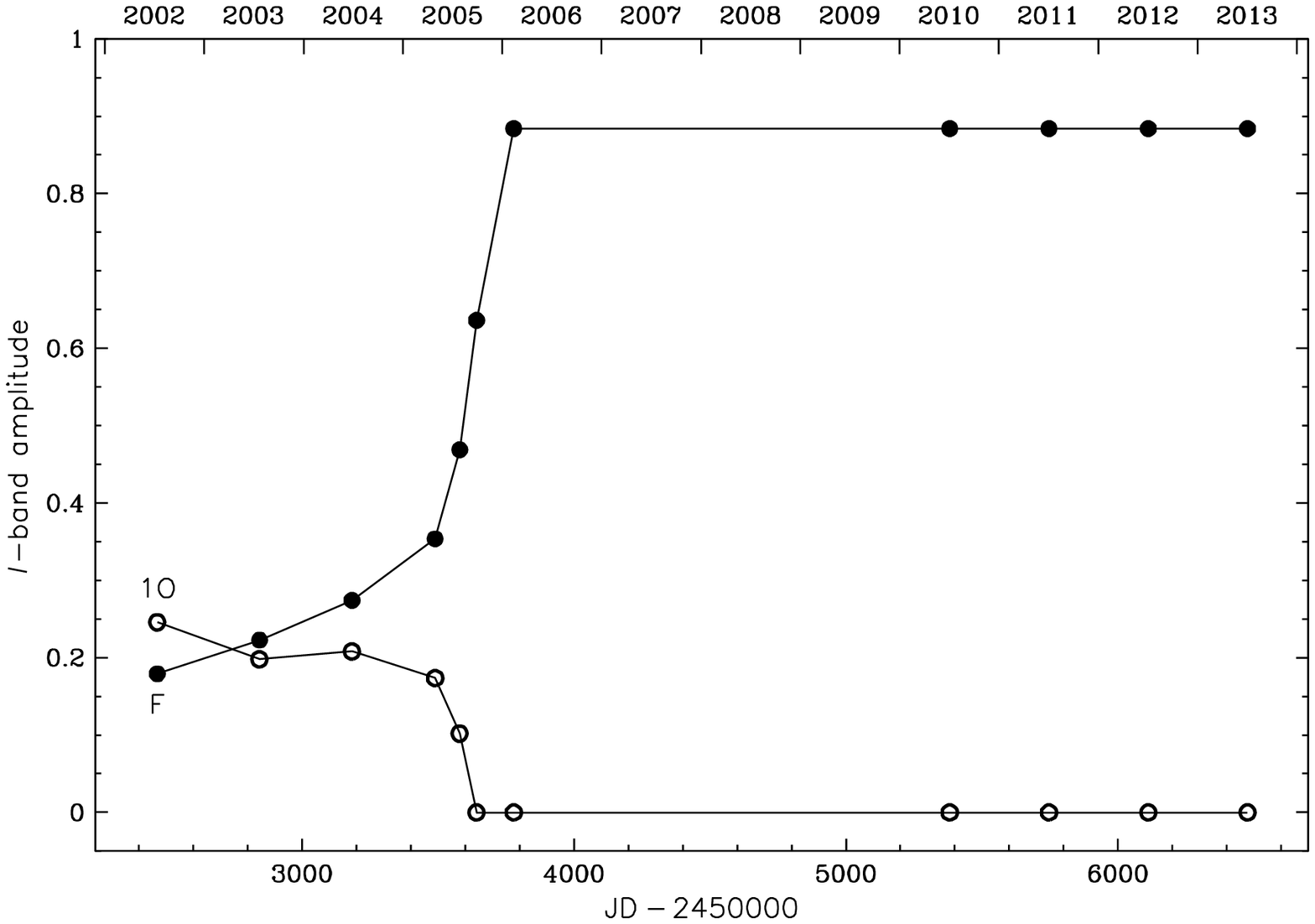}
\FigCap{Changes of the peak-to-peak {\it I}-band amplitudes of the
fundamental mode (filled circles) and first overtone (open circles) with
time.}
\end{figure}

In Fig.~4, we demonstrate the amplitude variations during the mode
switching. Unfortunately, in the 2006 season, when OGLE-BLG-RRLYR-12245
probably stabilized as a fundamental-mode pulsator, OGLE project collected
only two observing points at the beginning of the season (in
February). However, these two points can be phased with the RRab light
curve observed from 2010 by the OGLE-IV survey, assuming a pulsation period
of 0.409969~d -- only slightly shorter then the period observed today:
0.409975~d. Taking this into account, in Fig.~4 we assumed that the
amplitude of the fundamental mode reached the final value at the beginning
of 2006.

\vspace*{3pt}
\Section{Discussion}

OGLE-BLG-RRLYR-12245 is the second, after V79 in M3 (Clement \etal 1997),
RR Lyrae star where the transition from double- (F+1O) to mono- (F) mode
pulsation was observed, and the first one for which we had the opportunity
to follow in detail this process. The OGLE time-series data confirm
suggestions that changes in double-mode RR Lyr stars may occur in a
one-year time-span (e.g. Clement \etal 1997, Clementini \etal
2004). Clementini \etal (2004) mentioned a number of RRd stars showing change
of the dominant mode from usual 1O to F. Although we do not know (except for
V79) whether ultimately the 1O mode will disappear, it seems natural to
include all these objects together with our star in one class.

Following van Albada and Baker (1973) seminal idea, the type of RR~Lyr star
pulsation is being linked to evolution along the horizontal branch. Stars
evolving in the blueward direction in the H-R diagram pass successively
through the F-mode, either-or (EO), and 1O-mode domains of the instability
strip. Those evolving redward stars pass these domains in the reverse
order. In the EO domain, the choice of the pulsation mode depends on the
direction of evolution. The Osterhoff dichotomy is thought to arise from
this dependence. Although the cause of double mode pulsations is still
debated (see e.g. Szab\'o \etal 2004, Smolec and Moskalik 2010) there is
agreement that they occur in the EO domain. This also agrees with the
positions of RRd pulsators in the period--luminosity diagrams for
Magellanic Clouds (Soszyñski \etal 2009, 2010).

It is possible that the event observed in OGLE-BLG-RRLYR-12245 is connected
with crossing the EO/F boundary during the redward evolution. A significant
growth of the F-mode amplitude may occur with the rate similar to the
linear growth rate which may be faster than 1/d. However, the
$6\times10^{-5}$d/yr rate period increase, inferred from numbers given in
Fig.1, is by more than two orders of magnitude higher than the evolutionary
rate (see e.g. Le Borgne \etal 2007). The most likely cause of the period
increase is a nonlinear shift resulting from the increase of the pulsation
amplitude.

Clement \etal (1997) used the systematic growth of the F-mode and decline
1O-mode in two RRd stars in M3 as an evidence for the redward
evolution. However, the behavior of one of them (V79) turned out more
complicated. According to Goranskij \etal (2010), who made use of old
photographic data, the star was an F-mode pulsator between 1895 until
1992. Then, it switched quickly to F+1O pulsation, which lasted to 2007,
when the star became again a pure F-mode pulsator. Such a behavior cannot
be caused by consecutive crossings of the F/EO and EO/F boundaries. They
came too soon one after another. In the pure F-mode phases, pulsation in
this star exhibits fast changes. Before 1992, the period of V79 varied
significantly and for a short time in 1992 the star stopped to show any
detectable pulsation. In data taken after 2007, the Blazkho-type modulation
was found in V79. As we may see in Fig. 1, the pulsation amplitude in
OGLE-BLG-RRLYR-12245 does not show any changes in the F-mode phase.

Goranskij \etal (2010) blamed the peculiarities of V79 pulsation to
resonances between different radial modes. This is possible. There are
numerous examples of dynamical systems where resonances lead complicated
behaviors without external cause. Buchler and Koll\'ath (2011) suggested that
the 9:2 resonance between F- and 9O-modes is responsible for the Blazkho
effect. The possibility that the occurrence of RRd pulsators in the EO
domain is due to the $2\nu_{\rm1O}=\nu_{\rm F}+\nu_{\rm2O}$ has been
already mentioned by Soszyñski \etal (2011). The loci of this resonance in
the Petersen diagram are depicted in Fig.~2. We can see there that the
model of OGLE-BLG-RRLYR-12245 lies very close to the resonance center. We
do not know whether the switching between the RRd and RRab pulsation type
is due to this or any other resonance.

For our star, the interpretation of the mode switching as a consequence of
passing through the EO/F boundary during the evolution across the
instability strip cannot be excluded. However, it seems more likely that
this phenomenon results from the specific development of the instability in
the EO domain. In the similar manner the Blazkho effect may arise in the F
domain and fast period changes found in many RRc stars (\eg Jurcsik \etal
2001). Complex variations in time do not need external cause nor do they
require resonances. The Lorentz strange attractor is the most famous
example. However, looking for the necessary conditions for various forms of
pulsation remains a valid task. These conditions may help us explain why
only 0.5\% of the Galactic bulge RR Lyr stars are double-mode pulsators
(Soszyñski \etal 2011), while in other environments this percentage is much
higher. For example, in the Large and Small Magellanic Clouds RRd stars
represent, respectively, 4\% and 10\% of the total sample of RR Lyr stars
(Soszyñski \etal 2009, 2010). Maybe we will understand why in spite of much
larger number of RRc stars we have not yet seen the RRc/RRab mode
switching? Furthermore, if the $2/P_{\rm 1O}=1/P_{\rm F}+1/P_{\rm 2O}$
resonance is indeed the condition for double-mode pulsation we get a new
constrain on the star parameters.

\Acknow{This work has been supported by the Polish Ministry of Science and
Higher Education through the program ``Ideas Plus'' award No. IdP2012
000162. The research leading to these results has received funding from the
European Research Council under the European Community's Seventh Framework
Programme (FP7/2007-2013)/ERC grant agreement no. 246678. WAD has been supported
by Polish NCN grant DEC-2012/05/B/ST9/03932.}


\begin{references}
\refitem{Buchler, J.R., and Koll\'ath, Z.}{2011}{\ApJ}{731}{24}
\refitem{Clement, C.M., Ferance, S., and Simon, N.R.}{1993}{\ApJ}{412}{183}
\refitem{Clement,~C.M., Hilditch,~R.W., Kaluzny,~J., and Rucinski,~S.M.}{1997}{\ApJ}{489}{L55}
\refitem{Clement, C.M., and Goranskij, V.P.}{1999}{\ApJ}{513}{767} 
\refitem{Clementini, G., Corwin, T.M., Carney, B.W., and Sumerel, A.N.}{2004}{\AJ}{127}{938}
\refitem{Goranskij~V.P., Clement~C.M., and Thompson~M.}{2010}{~}{~}{in Variable Stars, the Galactic halo and Galaxy Formation, ed. Sterken~C., Samus~N. and~Szabados~L. (Moscow: Sternberg Astronomical Institute of Moscow Univ.), p.~115}
\refitem{Jurcsik, J., Clement, C., Geyer, E.H., and Domsa, I.}{2001}{\AJ}{121}{951}
\refitem{Kaluzny, J., Hilditch, R.W., Clement, C., and Rucinski,~S.M.}{1998}{\MNRAS}{296}{347}
\refitem{Le Borgne, J.F., Paschke, A., Vandenbroere, J., Poretti, E., Klotz, A., Bo{\"e}r, M., Damerdji, Y., Martignoni, M., and Acerbi, F.}{2007}{\AA}{476}{307} 
\refitem{Smolec, R., and Moskalik, P.}{2010}{\AA}{524}{A40} 
\refitem{Soszyñski,~I., Udalski, A., Szymañski,~M.K., Kubiak,~M., Pietrzyñski,~G., Wyrzykowski,~£., Szewczyk,~O., Ulaczyk,~K., and Poleski,~R.}{2009}{\Acta}{59}{1}
\refitem{Soszyñski,~I., Udalski, A., Szymañski,~M.K., Kubiak,~M., Pietrzyñski,~G., Wyrzykowski,~£., Ulaczyk,~K., and Poleski,~R.}{2010}{\Acta}{60}{165}
\refitem{Soszyñski,~I., Dziembowski, W.A., Udalski, A., Poleski,~R., Szymañski,~M.K., Kubiak,~M., Pietrzyñski,~G., Wyrzykowski,~£., Ulaczyk,~K., Koz³owski,~S., and Pietrukowicz,~P.}{2011}{\Acta}{61}{1}
\refitem{Szab{\'o}, R., Koll{\'a}th, Z., and Buchler, J.R.}{2004}{\AA}{425}{627}
\refitem{Udalski, A.}{2003}{\Acta}{53}{291}
\refitem{Udalski, A., Szymañski, M.K., Soszyñski, I., and Poleski, R.}{2008}{\Acta}{58}{69}
\refitem{van Albada, T.S., and Baker, N.}{1973}{\ApJ}{185}{477}
\end{references}
\end{document}